\documentclass[12pt,preprint]{aastex}

\def\simlt{\mathrel{\hbox{\rlap{\hbox{\lower4pt\hbox{$\sim$}}}\hbox{$<$}}}}
\def\simgt{\mathrel{\hbox{\rlap{\hbox{\lower4pt\hbox{$\sim$}}}\hbox{$>$}}}}

\newcommand{\perval}[2]{{#1\mbox{$^{#2}$}}}
\newcommand{\tee}[1]{\mbox{$\times 10^{#1}$}}
\newcommand{\persec}{\perval{s}{-1}\/}
\newcommand{\percm}{\mbox{$\cm^{-2}$}}
\newcommand{\cgsflux}{\erg~\percm~\persec}
\newcommand{\cgslum}{\erg~\persec}
\newcommand{\nh}{\mbox{$N_{\rm H}$}}
\newcommand\us{\mbox{ }\mu\mbox{s}}
\newcommand{\erg}{\mbox{$\rm\,erg$}\/}
\newcommand{\cm}{\mbox{$\rm\,cm$}}
\def\tuc {47~Tuc}
\newcommand{\chandra}{{\em Chandra\/}}

\newcommand{\psra}{B1821$-$24A}

\begin{document}
\title{Variability of Nineteen Millisecond Pulsars in 47 Tucanae with
{\em CHANDRA}/HRC-S}
\author{
P.~B.~Cameron\altaffilmark{1},
R.~E.~Rutledge\altaffilmark{2},
F.~Camilo\altaffilmark{3},
L.~Bildsten\altaffilmark{4},
S.~M.~Ransom\altaffilmark{5},
and S.~R.~Kulkarni\altaffilmark{1} 
}

\altaffiltext{1}{Division of Physics, Mathematics and Astronomy,
California Institute of Technology, MS 105-24, Pasadena, CA 91125;
pbc@astro.caltech.edu, srk@astro.caltech.edu}
\altaffiltext{2}{Department of Physics, McGill University,
Rutherford Physics Building, 3600 University Street, Montreal, QC H3A
2T8, Canada; rutledge@physics.mcgill.ca}
\altaffiltext{3}{Columbia Astrophysics Laboratory, Columbia
University, 550 West 120th Street, New York, NY 10027;
fernando@astro.columbia.edu}
\altaffiltext{4}{Kavli Institute for Theoretical Physics and
Department of Physics, Kohn Hall, University of California, Santa
Barbara, CA 93106; bildsten@kitp.ucsb.edu}
\altaffiltext{5}{National Radio Astronomy Observatory, 520 Edgemont
Road, Charlottesville, VA 22903; sransom@nrao.edu}

\begin{abstract}
We present results from our 830\,ksec observation of the globular
cluster 47~Tucanae with the \chandra\ {\em X-ray Observatory's} High
Resolution Camera-S.  We limit our analysis here to the 19 previously
known, localized millisecond pulsars (MSPs) in the cluster. This work
more than doubles the sample of X-ray-detected MSPs observed with
sensitivity to rotational variability; it is also the first survey of
a large group of radio-discovered MSPs for which no previous X-ray
pulsations have been detected and is therefore an unbiased survey of
the X-ray properties of radio-discovered MSPs.  We find that only
\tuc~D, O and R show significant pulsations at the $\simgt 4$-$\sigma$
level, but there is statistical evidence for rotational variability in
five additional MSPs.  Furthermore, we constrain the pulsed
magnetospheric emission of 7 more MSPs using Monte Carlo
simulations. The result is that the majority of the \tuc\ MSPs are
characterized by low pulsed fractions, $\simlt 50$\%.  In cases where
larger pulsed fractions are measured the folded pulse profiles show
relatively large duty cycles. When considered with previous
spectroscopic studies, this suggests that the X-ray emission arises
from the neutron star's heated polar caps, and in some cases, from
intra-binary shocks, but generally not directly from the star's
magnetosphere.  We discuss the impact of these results on our
understanding of high energy emission from MSPs.
\end{abstract}
\keywords{globular clusters: individual (47 Tucanae) --- pulsars:
general --- stars: neutron --- X-rays: stars}
 
\section{Introduction}
\label{sec:intro}
Millisecond pulsars (MSPs) are old neutron stars spun-up 
by accretion of mass and angular momentum from the matter of a donor
binary companion \citep{alpar82}. When compared to the canonical radio
pulsar population they are distinguished by short spin periods, $P
\simlt 25$\,ms, small spin-down rates, $\dot{P} \simgt 10^{-20}$\,s/s, and
thus low inferred dipole magnetic field strengths, $B_{\rm dipole}
\propto (P\dot{P})^{1/2} \sim 10^{8-10}$\,G, with large characteristic
ages, $\tau \equiv P/2\dot{P} \simgt 1$\,Gyr. Studies of the $\approx
150$ known MSPs are difficult at wavebands outside of the radio due to
their intrinsic faintness. The vast majority ($\approx 80$\%) of MSPs
have binary companions that dominate at optical wavelengths, thus
X-rays are an important avenue for studying MSPs. 

Currently, only 16 MSPs outside of 47 Tucanae (NGC 104, hereafter
\tuc) have been detected in X-rays, and only 12 of these have been
observed with sufficient time resolution to explore variability on
rotational timescales (see Table~\ref{tab:other}). There are several
proposed physical mechanisms capable of generating X-rays from these
MSPs. Non-thermal emission processes in the neutron star magnetosphere
generate power-law components in their X-ray spectra with
characteristic photon indices $\Gamma \approx$ 1.5--2. Pulsars in this
class (e.g. PSRs~B1937$+$21 and \psra; those above the first
horizontal line in Table~\ref{tab:other}) have large spin-down
energies ($\dot{E} \simgt 10^{35}$\,\cgslum), bright X-ray emission
($L_X \simgt 10^{32}$\,\cgslum), low duty cycles and pulse profiles that
closely resemble the radio emission in morphology and phase with large
pulsed fractions, $f_p \simgt 50$\% (see \S\ref{sec:rot}).  Power-law
spectral components can also be produced when the wind from the MSP
interacts with material from the binary companion causing an
intra-binary shock.  These pulsars have similar properties to those
above, with the exception that they lack strong rotational modulation
(e.g. the `black widow' pulsar, PSR~1957$+$20).  Finally, heating of
the neutron star polar cap by the bombardment of relativistic
particles provides a mechanism for producing thermal X-ray emission.
MSPs dominated by thermal spectra (e.g. PSRs~J0437$-$4715,
J2124$-$3358; most of those below the first horizontal line in
Table~\ref{tab:other}) are characterized by lower spin down energies
($\dot{E} \simlt 10^{34}$\,\cgslum), lower X-ray luminosities ($L_X
\simlt 10^{32}$\,\cgslum) and pulse profiles with large duty
cycles. As seen in Table~\ref{tab:other}, the pulsed fractions of
these MSPs are usually poorly constrained, but generally show $f_p
\approx 50$\%.  The emission of thermal cooling X-rays from the
neutron star surface and those from pulsar wind nebulae are not
thought to be important for these old objects, so we will not consider
them here.

The unprecedented spatial resolution of \chandra\ has enabled detailed
studies of MSPs in globular clusters.  Observations of M28, M4,
NGC~6397, M30 and others have provided the first census of
low-luminosity X-ray sources in these clusters
\citep{rutledge04,becker03,bassa04,grindlay02,ransom04}. However, the
largest endeavor has been {\em Chandra's} observing campaign of \tuc\
\citep{grindlay01, grindlay02, heinke05, bogdanov06}. This work has
shown that the spectral characteristics of the \tuc\ MSPs are
relatively homogeneous. Their luminosities fall in a narrow range,
$L_X \approx 10^{30-31}$\,\cgslum, and are well described by thermal
spectral models with small emission radii, $R_{\rm eff} \approx$
0.1-3\,km and temperatures of $T_{\rm eff} \approx
$1--3\tee{6}\,K. The only exceptions are the radio-eclipsing binaries
\tuc~J, O and W, which require additional power-law components above
2\,keV with $\Gamma \approx$ 1.0--1.5.  These results are reinforced
by the findings of detailed spectroscopic studies by {\em Chandra} and
{\em XMM-Newton} that have emphasized the dominant thermal components
in nearby MSP X-ray spectra over the fainter power-law features
\citep{zavlin02,zavlin06}. Consequently, we expect the predominant
X-ray emission from the MSPs in \tuc\ to arise from heated polar caps,
and to be modulated just by rotating a small area relative to the
observer (i.e. sinusoidal pulse profiles;
\citealt{grindlay02,cheng03,bogdanov06}).

In this paper, we present high time resolution data capable of
exploring the fast time variability of the X-ray counterparts to the
\tuc\ MSPs. In \S~\ref{sec:obs} we present the details of the
observations and data reduction, followed by the variability analysis
(\S~\ref{sec:var}) and an examination of the accuracy of the HRC-S
time tags (\S~\ref{sec:hrc}). We find that the HRC-S remains capable
of detecting fast modulation, but the MSPs in \tuc\ lack strong
variability on all timescales probed. We discuss the impact of this
result on our understanding of the X-ray emission from MSPs in
\S~\ref{sec:dis}.

\section{Observations and Data Reduction}
\label{sec:obs}
Observations of \tuc\ were performed with {\em Chandra's} HRC-S
detector \citep{hrc1,hrc2}. They began 2005 Dec 19 7:20 UT
with 14 subsequent visits over the next 20\,days for a total of
833.9\,ks of exposure time (see Table~\ref{tab:obs} for a
summary). The observing plan of dividing the 833.9 ksec into
50--100\,ksec visits spread out over 20 days was adopted -- instead
of the optimal choice for pulsar detection of an uninterrupted
observation -- to mitigate a thermal limitation in space-craft
operations. The HRC-S has a timing resolution of 15.625\,$\us$ in the
nominal energy range of 0.1-2\,keV, although it has essentially no
energy resolution.  The data were analyzed using Chandra Interactive
Analysis of Observations\footnote{http://cxc.harvard.edu/ciao/}
software (CIAO) version 3.3 and CALDB version 3.2.1.

We began the data analysis of each observation by registering it to
the first pointing (ObsID 5542) using the relative astrometry of the
four brightest sources in the field. These corrections typically
resulted in $\simlt$ 0\farcs5 corrections to the native astrometry.
Data were filtered on pulse invariant (PI) channel to maximize the
signal-to-noise ratio (SNR) for the known MSPs using the following
approach.  We compared the PI channel distribution of all counts
extracted from regions corresponding to the known MSPs with that
extracted from a background region.  The background and source PI
distributions are identical below PI $=$ 25 (within 1\% in
counts/area).  Above this value the source region counts are
significantly in excess, so we adopt PI $=$ 25 as the lower PI limit
for background filtering.  We determined the upper-limit in PI channel
by maximizing the total source SNR, finding the maximum when data at
PI $>$ 120 are excluded from analysis.  Thus, we adopt a PI range of
25--120 for all analyses; this decreases the MSP source count rates by
6\%, while excluding 51\% of the total background counts.  The
resulting background contribution is 17.9 counts\,arcsec$^{-2}$ over
the entire observing span. 

Currently, positions are known for 19 of the 22 MSPs in \tuc.
Published pulsar timing solutions can account for 16 of these
positions \citep{freire01,freire03}, while \tuc~R and Y have
unpublished solutions (Freire et al. in preparation).  \tuc~W has only
a preliminary timing solution, but was localized by the eclipses of
its optical counterpart (Freire et al. unpublished;
\citealt{edmonds02}). The close proximity of pulsars \tuc~G and I and
\tuc~F and S (separations 0\farcs12 and 0\farcs7, respectively) do not
allow them to be resolved by \chandra, so their counts must be
considered jointly. For analysis purposes we attribute 50\% of
detected photons to each pulsar (see \S\ref{sec:rot}).

We extracted photons from within circular regions surrounding each of
the known MSP positions. The size of each extraction region can be
found in Table~\ref{tab:money}. The radius was chosen adaptively in
order to maximize SNR, but constrained to mitigate contamination by
nearby objects. However, some contamination due to source crowding is
unavoidable. We estimate this contribution by modeling the PSF as a
Gaussian with $\sigma = 0.29$\arcsec, calculating the number of
photons that fall in the extraction region of a given source from each
neighboring source, and using this estimate to update the extracted
source counts. We iterate this procedure until we have an estimate of
the source crowding contamination for all sources in the field. This
analysis shows that the contamination is negligible ($<$ 1\% of
extracted counts) for all MSPs except O and R, which each have $\sim
13$ additional background counts due to nearby sources. We include
these in our estimate of their backgrounds in the subsequent analysis
(Table~\ref{tab:money}). Assuming these background estimates, we
detect sources at each of the 17 independent pulsar positions with $>
5$-$\sigma$ significance.

Throughout the analysis we assume a distance of 4.85\,kpc to \tuc\
\citep{gratton03}.  We apply an energy correction factor of
5.044\tee{-12}\,\cgsflux\ or 1.42\tee{34}\,\cgslum\ (0.5-2.0\,keV) per 1
HRC-S count/sec, which is the unabsorbed X-ray flux assuming
\nh=1.3\tee{20}\,\percm\ and blackbody emission with a temperature of
0.178\,keV determined from
WebPIMMS\footnote{http://heasarc.gsfc.nasa.gov/Tools/w3pimms.html}
\citep{bogdanov06}.  In addition to the extraction region size, we
list the total counts extracted, expected background counts and the
time averaged luminosity in Table~\ref{tab:money}.

\section{Variability Analysis}
\label{sec:var}
Prior to the timing analysis, we use the CIAO tool {\tt axbary} to
convert the event times to the solar system barycenter using the
JPL~DE200 solar system ephemeris, the \chandra\ orbital ephemeris and
the radio/optical position of each MSP (see \S~\ref{sec:obs}). 

\subsection{Rotational Variability}
\label{sec:rot}
We calculated the rotational phase of each arriving photon for each
MSP using the latest radio ephemerides corrected to X-ray frequencies
(\citealt{freire03}; Freire et al., in preparation). The resulting
phases were searched for variability with the $Z_n^2$-test
\citep{buccheri83}, where $n$ is the optimal number of harmonics as
determined from the H-test (\citealt{dejager89}). The variable $Z_n^2$
has a probability density function distributed as $\chi^2$ with 2$n$
degrees of freedom. We list the value of this variable, the detection
significance in equivalent Gaussian $\sigma$ and $n$ in
Table~\ref{tab:money}.  Only \tuc~D, O and R show variability above
the $\approx$ 4-$\sigma$ level. Their folded pulse profiles are shown
in Figure~\ref{fig:rot}. All three pulsars' profiles are characterized
by large duty cycles. \tuc~O shows evidence for two peaks centered at
phases, $\phi \approx 0.0$ and $\phi \approx 0.4$ with widths of
$\delta\phi \approx 0.2$ and $\delta\phi \approx 0.4$,
respectively. Only a single peak is evident in the profiles of \tuc~D
and R centered at $\phi \approx 0.45$ with $\delta\phi \approx 0.25$
and $\phi \approx 0.5$ with $\delta\phi \approx 0.3$, respectively.

As seen in Table~\ref{tab:money}, 5 of the 19 MSPs are detected with
marginal significance, 2.8--3.3-$\sigma$. Given the relatively large
size of our MSP sample, we can quantify the significance of these
marginal detections. We are free to choose the significance level with
which we call an MSP `variable'.  Once we choose this level, the
problem becomes one of binomial statistics where each MSP represents
an independent measurement for variability.  Choosing the 99\% percent
confidence level for $Z^2_n$ (corresponding to 2.58-$\sigma$) allows
us to identify 8 `variable' sources when only $\approx$ 0.2 are
expected if the MSPs were drawn from a random distribution.  The
probability of 8 of 19 trials being labeled `variable' is
6.7\tee{-12}. The probability of one or more of these being false
detections is 17.4\%.  Since the confidence level at which we label an
MSP as `variable' is arbitrary, we list a range of confidence levels
and the number of corresponding detections with the binomial and false
detection probabilities in Table~\ref{tab:prob}. For the remainder of
our analysis we will adopt the 99\% confidence level.  This results in
\tuc~D, E, F, H, O, Q, R and S being labeled as `variable.' Their
profiles are shown in Figure~\ref{fig:rot}.

The pulsed fraction, $f_p$, for each MSP can be determined using two
steps. First, we estimate the DC (unpulsed) level with the
non-parametric bootstrapping method of \cite{swanepoel96}.  The
advantage of this technique is that it works on the raw phases without
the need to construct a phase histogram or know the pulse shape a
priori. This level is shown by the solid line in Figure~\ref{fig:rot}
with $\pm 1$-$\sigma$ errors. Next, we correct for the fact that the
DC component includes both unpulsed photons from the MSP and
background photons. Note that for MSPs F and S we consider 50\% of the
background subtracted source counts to be unpulsed background
photons. The expected background contribution to the pulse profile is
denoted by the dashed line in Figure~\ref{fig:rot} (see
\S~\ref{sec:obs}). The pulsed fraction determined by the bootstrapping
method can be related to the true pulsed fraction by $f_p = f_{p,{\rm
boot}} N_t/(N_t - N_b)$, where $f_{p,{\rm boot}}$ is the pulsed
fraction determined by the bootstrapping method, $N_t$ is the total
number of extracted counts and $N_b$ is the total number of background
counts contributing to $N_t$.  Both $N_t$ and $N_b$ can be found in
Table~\ref{tab:money}. The pulsed fractions derived in this manner are
also listed in Table~\ref{tab:money}.

We calculate upper limits on the pulsed fractions of the remaining 11
MSPs with timing solutions (note that W has only a preliminary
solution) with Monte Carlo simulations assuming two scenarios. First,
we make the conservative assumption that the underlying pulse shape is
sinusoidal. For each MSP and pulsed fraction we simulate 500 light
curves with a total number of counts $N_t = N_b + N_s$, where $N_s$ is
the number of source counts. If the pulsed fraction is $f_{p,{\rm
sine}}$, then $N_s$ consists of $(1-f_{p,{\rm sine}})N_s$ unpulsed
counts and $f_{p,{\rm sine}}N_s$ pulsed counts. For the two unresolved
MSPs that do not show pulsations (G and I) we assume that 50\% of the
source photons constitute an unpulsed background. We determine the
90\% confidence upper limit on the pulsed fraction as the value of
$f_{p,{\rm sine}}$ at which 450 synthetic light curves have values of
$Z^2_1 \simgt 27.4$ (corresponding to 5-$\sigma$). The result is that
only J, L, W and Y have sufficient counts to be constrained in this
way.
 
Motivated by the close correspondence of the radio and X-ray pulse
profiles of magnetospherically dominated MSPs (see \S~\ref{sec:intro}),
we determine a second set of upper limits based on the assumption that
the underlying X-ray pulse has the same morphology as the radio pulse.
We denote this pulsed fraction as $f_{p,{\rm radio}}$.  In the case of the
weak radio pulsars, N, R, T, W and Y it was necessary to model the
radio pulse(s) with Gaussians and use these as the assumed X-ray pulse
shape. The radio pulse profiles for remaining MSPs have sufficient
signal-to-noise to be used directly.  We then perform the Monte Carlo
simulations with the technique described above to determine the 90\%
confidence upper limits.  The results can be found in
Table~\ref{tab:money}. It is possible to constrain the X-ray emission
with the same morphology as the radio pulse from 7 pulsars in this
way.

\subsection{Orbital Variability}
The orbital periods of the 12 binary \tuc\ MSPs span the range
0.07--2.4\,days. In order to search for variability during these
orbits we calculated the orbital phase of each arriving photon for
each MSP and constructed histograms with 5, 10 and 20 bins. We then
corrected each bin for the variation of exposure time during that
particular orbital phase so that we have a histogram of counts per
second per bin and subtracted the expected background contribution. We
search for variability by computing $\chi^2$ between the histogram and
a constant count rate. We found that none of the \tuc\ binary MSPs
showed significant orbital variability.

\begin{figure}
\epsscale{.8}
\plotone{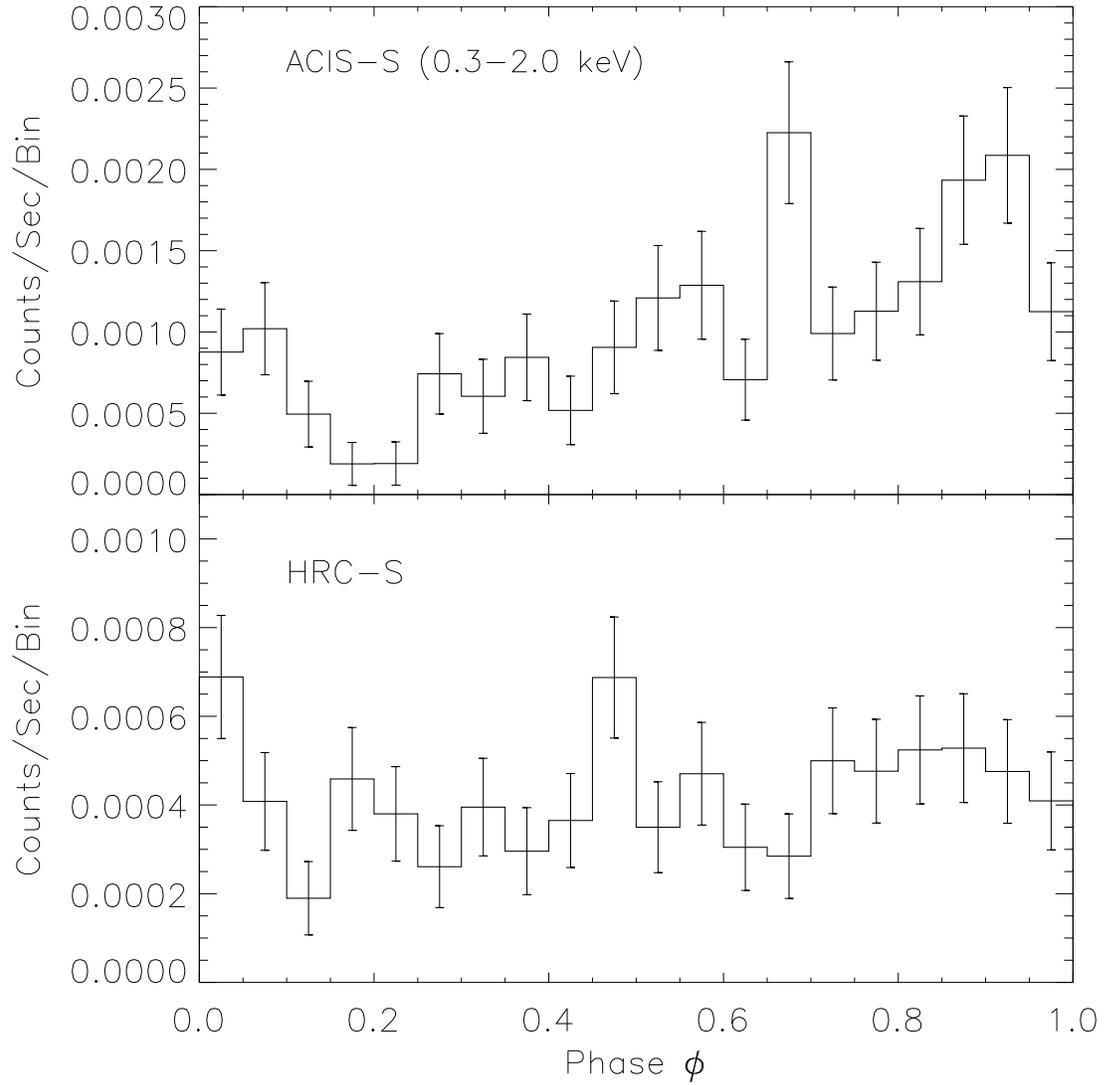}
\caption{Orbital Profiles of \tuc~W from ACIS-S (upper) and HRC-S
(lower).  The error bars are 1-$\sigma$ and histograms contain 20 bins
per period.}
\label{fig:mspw}
\end{figure}

Substantial X-ray eclipses characterized by a complete disappearance
of hard ($> 2$\,keV) photons and a decline in soft ($< 2$\,keV)
photons for $\approx$ 30\% of the orbit have been reported for \tuc~W
\citep{bogdanov05}. Thus, the lack orbital variability in \tuc~W is
surprising. We reduced archival ACIS-S data of \tuc~W from ObsIDs
2735, 2736, 2737, and 2739 in order to quantify the properties of the
eclipse in the HRC-S bandpass. We reprocessed the level 1 event files
to make use of the latest calibration and filtered periods of high
background flaring (i.e. periods with count rates $>$ 3-$\sigma$ above
the mean full-frame rate). This resulted in 251.3\,ksec of exposure
time. We computed the orbital phase of each photon within a 1\arcsec\
radius of the position of \tuc~W in the energy range 0.3--2.0\,keV
with the preliminary timing solution (see \S~\ref{sec:obs}).  The
X-ray eclipse is evident in the histogram shown in
Figure~\ref{fig:mspw}a at phase, $\phi \approx$ 0.2, where $\phi=0$ at
the time of the ascending node.  This X-ray eclipse timing agrees well
with observed radio eclipse from $\phi =$ 0.1--0.4
\citep{camilo00,edmonds02}. The variability is significant at the 4.3-$\sigma$
level from the measured value of $Z^2_1=30.1$ and we determine an
orbital modulation of $f_p = 36 \pm 9$\% using the non-parametric
bootstrap method (see \S~\ref{sec:rot}). This is consistent with the
90\% confidence limit of $f_{p,{\rm sine}} < 48$\% on a sinusoidal
signal in the HRC-S time series of \tuc~W (see \S~\ref{sec:rot}).

\subsection{Aperiodic Variability}
To search for aperiodic variability in the \tuc\ MSPs, we have applied
the Bayesian blocks algorithm of \cite{scargle98} as implemented by
the Interactive Spectral Interpretation System (ISIS;
\citealt{houck00}). The algorithm determines the optimal decomposition
of the light curve into constant count rate segments based on a
parametric maximum likelihood model of a Poisson process. The raw
(unbinned) events are divided into `blocks' and the odds ratio that
the count rate has varied is computed. If variability is found, each
`block' is further subdivided to characterize the structure of the
light curve (e.g. step-function variation, flaring, etc.).  We could
not identify any intra-observation variability from 17 MSPs using an
odds ratio corresponding to 68\% chance that any variability is real.
In addition, the inter-observation count rates for each MSP derived
from this process do not show significant variability over the
$\approx$ 20\,day span of our observations. Thus, the X-ray emission
from the \tuc\ MSPs is stable on timescales ranging from minutes to
days.

\subsection{ACIS-S vs. HRC-S Comparison}
\begin{figure}
\plotone{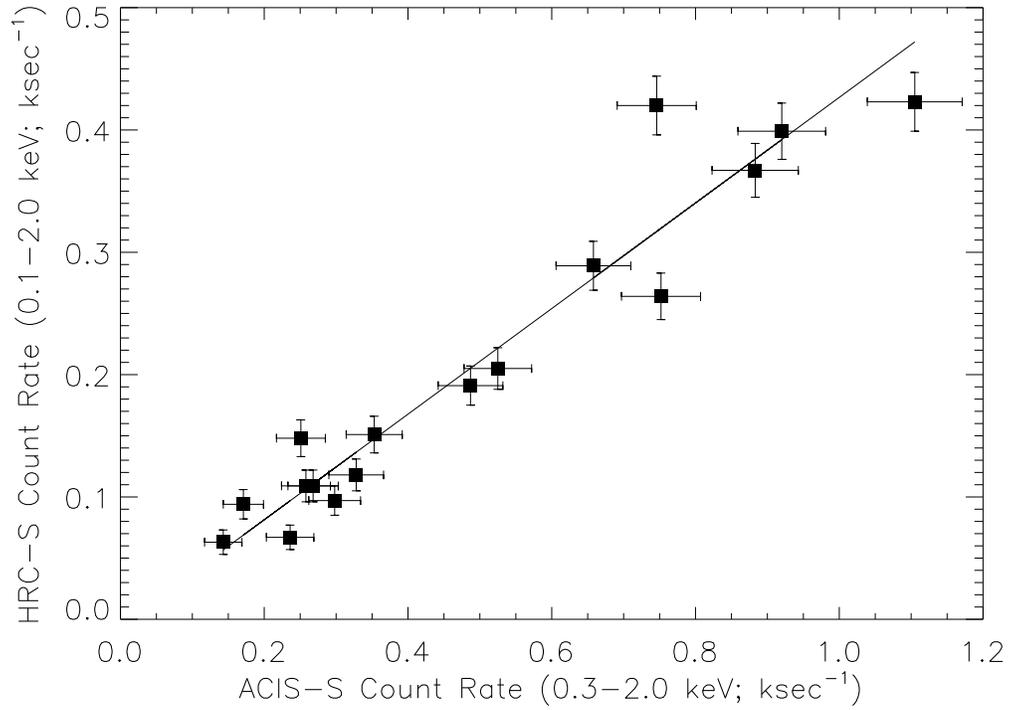}
\caption{HRC-S/ACIS-S count rate comparison using the 17 independent
\tuc\ MSP detections. The line represents the best linear fit to the
data.}
\label{fig:compare}
\end{figure}

With such a large sample of constant luminosity X-ray sources, we can
compare the count rates between the ACIS-S and HRC-S for soft thermal
sources.  In Figure~\ref{fig:compare} we compare the count rate for
the two detectors using each of the 17 independent MSP detections. For
the ACIS-S count rate, we summed the values in the 0.3--0.8\,keV and
0.8--2.0\,keV bands listed by Heinke et al. (2005).  The relation
between count rates in the two different detectors was $I({\rm HRC-S})
= (0.43\pm0.024)\times I({\rm ACIS-S})$. This is consistent with the
conversion from HRC-S to ACIS-I \citep{rutledge04}.

\section{HRC-S Timing Accuracy}
\label{sec:hrc}
The accuracy of the HRC-S time tags was demonstrated to be $\pm 12
\us$ in an observation of M28 \citep{rutledge04}. We have investigated
several issues to ensure that the HRC-S has sufficient accuracy to
detect MSPs. In addition to accounting for a leap second that occurred
during the middle of the observing span, we examined the effect of
telemetry saturation, different solar system ephemerides and analyzed
a recent calibration observation of the globular cluster M28.

\subsection{Telemetry Saturation}
The maximum telemetered full-field, unfiltered count rate for the HRC-S
is 184 counts \persec (\chandra\ Proposers' Observatory Guide
\footnote{http://cxc.harvard.edu/proposer/POG/index.html}).  At rates
above this, a decreasing fraction of all counts will be telemetered.
The data will be affected if this count rate is exceed during the
2.05\,s full-frame readout time.  The effect of telemetry saturation
on timing certainty is that, when telemetry is saturated, not all
events are telemetered back to Earth.  Due to the HRC wiring error
\citep{tennant01}, the $N$th event detected by the HRC-S has its time
assigned to the $N$+1st event\footnotemark[8].  The true time series
can be reconstructed if both are telemetered; if either is not, then
the wrong time will be assigned to one event (either the $N$th event,
or the $N-1$st event).

To test the extent to which this saturation impacts our observations
we constructed a light curve of our entire dataset binned in 2.05\,s
intervals. We found that only $\approx 0.1$\% of the bins exceeded the
maximum count rate. Thus telemetry saturation will effect only
$\approx 0.1$\% of the counts in any given MSP, which is negligible.

\subsection{Ephemeris Comparison}
We performed a preliminary extraction of data using DE405 and compared
the timing precision of data corrected to DE200 using one of the
brightest sources in the field.  We found the entire observational
period was offset by $t_{\rm DE405}-t_{\rm DE200} = 1.3809$\,ms at the
start of the observational period, decreasing monotonically to
1.3719\,ms at the end of the observation period.  Thus, there was an
average direct offset between the photon time-of-arrivals (TOAs) in
the two ephemerides of $\approx$1.3764\,ms, and a range of variation
of $\approx 9\, \mu s$.  This $9\, \mu s$ therefore amounts to the
relative timing uncertainty due to the adopted ephemeris, comparable
to the uncertainties in HRC digitization ($\pm 5\,\mu s$) and Chandra
clock stability ($\pm 5\,\mu s$) \citep{rutledge04}.

\subsection{s/c Clock Stability}
A calibration observation of PSR~\psra\ was performed on May 27,
2006 starting at 12:30\,UT for 40887\,sec in order to evaluate the
stability of the HRC-S clock. A complete analysis is beyond the scope
of this work. However, a search of the data using the known radio
timing ephemeris \citep{rutledge04} clearly detects the 3.05\,ms
pulsar with $Z^2_1 = 330$ corresponding to a detection significance of
$\approx 18$-$\sigma$. Thus, we conclude the HRC-S clock has remained
sufficiently stable to detect MSPs.

\section{Discussion and Conclusions}
\label{sec:dis}
\begin{figure}      
\plotone{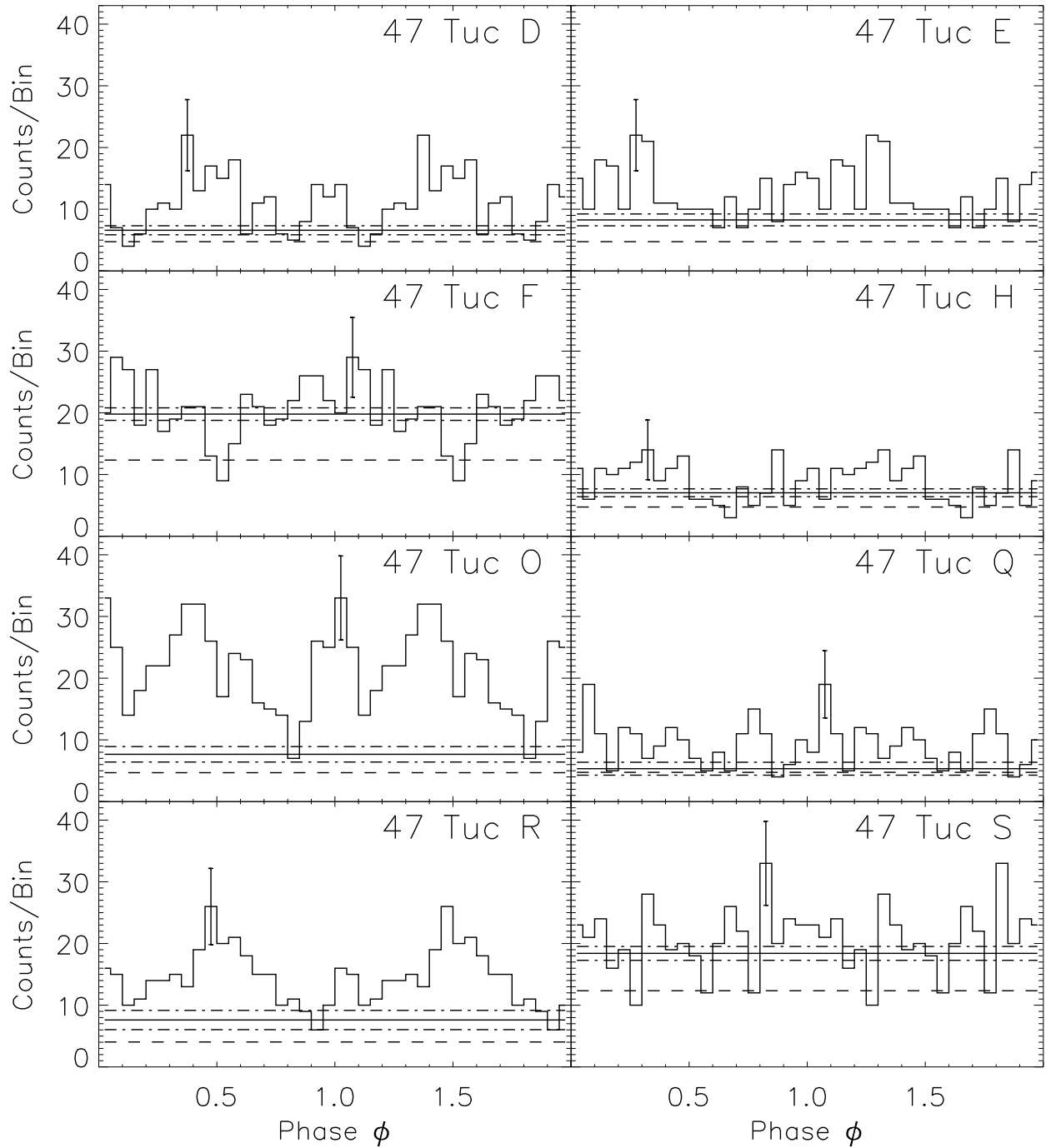}
\bigskip
\caption{Pulse profiles of the variable \tuc\ MSPs with a typical
error bar (two periods are shown for clarity). The histograms were
constructed with 20 bins per period. The solid horizontal line denotes
the DC (unpulsed) contribution to the pulse profile as determined by
the nonparametric bootstrapping algorithm with the associated
1-$\sigma$ uncertainties denoted with dashed-dotted lines.  The dashed
line denotes the estimated contribution to this level due to the
uniform background and source crowding (50\% of the background
subtracted counts are attributed to this total for MSPs F and S; see
\S~\ref{sec:rot} for details). }
\label{fig:rot}
\end{figure}

\chandra\ HRC-S observations of \tuc\ have allowed us to study a
relatively large sample of 19 MSPs on millisecond to week timescales.
We find that the MSPs in \tuc\ uniformly show very low levels of
variability on all scales probed.  We have sufficient statistics to
meaningfully constrain (under the assumption that their X-ray pulse
profiles match the radio) or measure the rotational modulation of 15
MSPs.  Eight of these objects have low pulsed fractions, $f_p \simlt
50$\%.  MSPs \tuc~D, O, and R each have pulsations detected at
$\simgt$ 4-$\sigma$ significance with relatively large pulsed
fractions, $f_p \simgt 60$\%, which are similar to the levels seen
from luminous MSPs dominated by non-thermal emission (e.g.
PSRs~\psra\ and B1937$+$21). However, the pulse profiles of these
objects (see Figure~\ref{fig:rot}) are characterized by large
duty-cycle features that do not resemble the sharp, low duty-cycle
profiles seen in the non-thermal MSPs (e.g. see \citealt{becker02}).

The existing ACIS data show that the \tuc\ MSPs have fairly
homogeneous spectroscopic properties
\citep{grindlay02,bogdanov06}. All but 3 of the \tuc\ MSPs are
characterized by 1--3\tee{6}\,K thermal spectra with low luminosities
in a narrow range, $L_{\rm X} \approx 10^{30-31}$\,\cgslum, and have
small emission radii, $R_{\rm eff} \approx$ 0.1--3.0\,km.  The low
level of measured variability presented here indicates that rotational
averaging does little to affect these values, which agree with the
predictions of polar cap heating scenarios \citep{harding02}. Thus we
conclude that, unlike the luminous non-thermal MSPs, the vast majority
of the X-ray emission from the \tuc\ MSPs is created by the heating of
the neutron star polar cap by a return current of relativistic
particles produced in the magnetosphere
\citep{arons81,harding01,harding02,grindlay02,bogdanov06}. For older
MSPs with very short spin periods and low magnetic fields, like those
in \tuc, the main source of the $e^{\pm}$-pair production is thought
to be through inverse Compton scattering of thermal X-rays from the
neutron star surface off of electrons in the pulsar magnetosphere
\citep{harding02}.

Only the radio-eclipsing binaries \tuc~J, O and W show power-law
spectral components that contribute significantly (70\%, 50\% and
75\%, respectively) to their total flux \citep{bogdanov06}. The lack
of strongly pulsed emission in \tuc~J and W suggests that the X-ray
emission does not arise in the neutron star magnetosphere, but instead
is likely the consequence of an intra-binary shock. This is in
agreement with conclusions based on orbital phase resolved
spectroscopy of \tuc~W by \cite{bogdanov05}. Conversely, the current
data do not conclusively identify the origin of X-rays from \tuc~O,
which shows significant pulsations. The X-rays from an intra-binary
shock are not expected to be modulated at the rotational period of the
MSP, so the measured pulsed fraction, $f_p = 83 \pm 21$\%, is only
marginally consistent with the 50\% spectroscopic allocation of X-rays
due to a shock. In addition, its large duty-cycle does not immediately
imply that polar cap heating is the source of the pulsed X-rays, since
broadly beamed magnetospheric emission viewed off-axis would appear to
have a large duty-cycle \citep{becker99}.  

The apparent non-detection of low duty-cycle pulsars is significant in
comparison with the pulse profile of \psra.  If all 19 MSPs had X-ray
pulse profiles identical to that of PSR~\psra, all would have been
detected with $\simgt$ 7-$\sigma$ significance (which we find for the
lowest SNR MSP, \tuc~T). Those MSPs with higher count rates would have
been detected with even greater significance. The implication is that
PSR~\psra\ has an unusually low duty cycle for a MSP.  If we assume
that low duty-cycle MSPs make up a fraction $f$ of the globular
cluster population, then the non-detection of even 1 such pulsar in 47
Tuc implies that \psra-like pulsars comprise $f<$20\% of the MSP
population in GCs (90\% confidence limit).  This limit could be much
lower, if the intrinsic distribution of duty cycles in magnetospheric
pulsars is lower than that of \psra, for example; however, there seems
to be little observational work quantifying the distribution of duty
cycles of observed pulsars.

\acknowledgements 
The authors would like to thank P. Freire for providing us with data
prior to publication.  Support for this work was provided by NASA
through Chandra Award Number G05-6060 issued by the Chandra X-ray
Observatory Center, which is operated by the Smithsonian Astrophysical
Observatory for and on behalf of NASA under contract NAS8-03060.
L.~B. acknowledges support from the NSF through grant PHY99-07949.

{\it Facilities:} \facility{CXO (HRC)}

\bibliographystyle{apj}

\begin{thebibliography}{43}
\expandafter\ifx\csname natexlab\endcsname\relax\def\natexlab#1{#1}\fi

\bibitem[{{Alpar} {et~al.}(1982){Alpar}, {Cheng}, {Ruderman}, \&
  {Shaham}}]{alpar82}
{Alpar}, M.~A., {Cheng}, A.~F., {Ruderman}, M.~A., \& {Shaham}, J. 1982, \nat,
  300, 728

\bibitem[{{Arons}(1981)}]{arons81}
{Arons}, J. 1981, \apj, 248, 1099

\bibitem[{{Bassa} {et~al.}(2004){Bassa}, {Pooley}, {Homer}, {Verbunt},
  {Gaensler}, {Lewin}, {Anderson}, {Margon}, {Kaspi}, \& {van der
  Klis}}]{bassa04}
{Bassa}, C., {Pooley}, D., {Homer}, L., {Verbunt}, F., {Gaensler}, B.~M.,
  {Lewin}, W.~H.~G., {Anderson}, S.~F., {Margon}, B., {Kaspi}, V.~M., \& {van
  der Klis}, M. 2004, \apj, 609, 755

\bibitem[Becker \& Tr{\"u}mper(1999)]{bt99} Becker, W., \& 
Tr{\"u}mper, J.\ 1999, \aap, 341, 803 

\bibitem[{{Becker} \& {Aschenbach}(2002)}]{becker02}
{Becker}, W. \& {Aschenbach}, B. 2002, in Neutron Stars, Pulsars, and Supernova
  Remnants, ed. W. Becker, H. Lesch, and J. Trümper (MPE Rep. 278; Carching:
  MPE), 64

\bibitem[{{Becker} {et~al.}(2003){Becker}, {Swartz}, {Pavlov}, {Elsner},
  {Grindlay}, {Mignani}, {Tennant}, {Backer}, {Pulone}, {Testa}, \&
  {Weisskopf}}]{becker03}
{Becker}, W., {Swartz}, D.~A., {Pavlov}, G.~G., {Elsner}, R.~F., {Grindlay},
  J., {Mignani}, R., {Tennant}, A.~F., {Backer}, D., {Pulone}, L., {Testa}, V.,
  \& {Weisskopf}, M.~C. 2003, \apj, 594, 798

\bibitem[{{Becker} \& {Tr{\"u}mper}(1999)}]{becker99}
{Becker}, W. \& {Tr{\"u}mper}, J. 1999, \aap, 341, 803

\bibitem[Blandford \& Teukolsky(1976)]{blandford76} Blandford, R., 
\& Teukolsky, S.~A.\ 1976, \apj, 205, 580 

\bibitem[{{Bogdanov} {et~al.}(2006){Bogdanov}, {Grindlay}, {Heinke}, {Camilo},
  {Freire}, \& {Becker}}]{bogdanov06}
{Bogdanov}, S., {Grindlay}, J.~E., {Heinke}, C.~O., {Camilo}, F., {Freire},
  P.~C.~C., \& {Becker}, W. 2006, \apj, 646, 1104

\bibitem[{{Bogdanov} {et~al.}(2005){Bogdanov}, {Grindlay}, \& {van den
  Berg}}]{bogdanov05}
{Bogdanov}, S., {Grindlay}, J.~E., \& {van den Berg}, M. 2005, \apj, 630, 1029

\bibitem[{{Buccheri} {et~al.}(1983){Buccheri}, {Bennett}, {Bignami}, {Bloemen},
  {Boriakoff}, {Caraveo}, {Hermsen}, {Kanbach}, {Manchester}, {Masnou},
  {Mayer-Hasselwander}, {Ozel}, {Paul}, {Sacco}, {Scarsi}, \&
  {Strong}}]{buccheri83}
{Buccheri}, R., {Bennett}, K., {Bignami}, G.~F., {Bloemen}, J.~B.~G.~M.,
  {Boriakoff}, V., {Caraveo}, P.~A., {Hermsen}, W., {Kanbach}, G.,
  {Manchester}, R.~N., {Masnou}, J.~L., {Mayer-Hasselwander}, H.~A., {Ozel},
  M.~E., {Paul}, J.~A., {Sacco}, B., {Scarsi}, L., \& {Strong}, A.~W. 1983,
  \aap, 128, 245

\bibitem[Camilo et al.(2000)]{camilo00} Camilo, F., Lorimer, 
D.~R., Freire, P., Lyne, A.~G., \& Manchester, R.~N.\ 2000, \apj, 535, 975 

\bibitem[{{Cheng} \& {Taam}(2003)}]{cheng03}
{Cheng}, K.~S. \& {Taam}, R.~E. 2003, \apj, 598, 1207

\bibitem[{{Cordes} \& {Lazio}(2002)}]{cordes02}
{Cordes}, J.~M. \& {Lazio}, T.~J.~W. 2002, preprint (astro-ph/0207156)

\bibitem[{{D'Amico} {et~al.}(2002){D'Amico}, {Possenti}, {Fici}, {Manchester},
  {Lyne}, {Camilo}, \& {Sarkissian}}]{damico02}
{D'Amico}, N., {Possenti}, A., {Fici}, L., {Manchester}, R.~N., {Lyne}, A.~G.,
  {Camilo}, F., \& {Sarkissian}, J. 2002, \apjl, 570, L89

\bibitem[{{D'Amico} {et~al.}(2001){D'Amico}, {Possenti}, {Manchester},
  {Sarkissian}, {Lyne}, \& {Camilo}}]{damico01}
{D'Amico}, N., {Possenti}, A., {Manchester}, R.~N., {Sarkissian}, J., {Lyne},
  A.~G., \& {Camilo}, F. 2001, \apjl, 561, L89

\bibitem[{{de Jager} {et~al.}(1989){de Jager}, {Raubenheimer}, \&
  {Swanepoel}}]{dejager89}
{de Jager}, O.~C., {Raubenheimer}, B.~C., \& {Swanepoel}, J.~W.~H. 1989, \aap,
  221, 180

\bibitem[{{Edmonds} {et~al.}(2002){Edmonds}, {Gilliland}, {Camilo}, {Heinke},
  \& {Grindlay}}]{edmonds02}
{Edmonds}, P.~D., {Gilliland}, R.~L., {Camilo}, F., {Heinke}, C.~O., \&
  {Grindlay}, J.~E. 2002, \apj, 579, 741

\bibitem[{{Freire} {et~al.}(2003){Freire}, {Camilo}, {Kramer}, {Lorimer},
  {Lyne}, {Manchester}, \& {D'Amico}}]{freire03}
{Freire}, P.~C., {Camilo}, F., {Kramer}, M., {Lorimer}, D.~R., {Lyne}, A.~G.,
  {Manchester}, R.~N., \& {D'Amico}, N. 2003, \mnras, 340, 1359

\bibitem[{{Freire} {et~al.}(2001){Freire}, {Camilo}, {Lorimer}, {Lyne},
  {Manchester}, \& {D'Amico}}]{freire01}
{Freire}, P.~C., {Camilo}, F., {Lorimer}, D.~R., {Lyne}, A.~G., {Manchester},
  R.~N., \& {D'Amico}, N. 2001, \mnras, 326, 901

\bibitem[{{Gratton} {et~al.}(2003){Gratton}, {Bragaglia}, {Carretta},
  {Clementini}, {Desidera}, {Grundahl}, \& {Lucatello}}]{gratton03}
{Gratton}, R.~G., {Bragaglia}, A., {Carretta}, E., {Clementini}, G.,
  {Desidera}, S., {Grundahl}, F., \& {Lucatello}, S. 2003, \aap, 408, 529

\bibitem[{{Grindlay} {et~al.}(2002){Grindlay}, {Camilo}, {Heinke}, {Edmonds},
  {Cohn}, \& {Lugger}}]{grindlay02}
{Grindlay}, J.~E., {Camilo}, F., {Heinke}, C.~O., {Edmonds}, P.~D., {Cohn}, H.,
  \& {Lugger}, P. 2002, \apj, 581, 470

\bibitem[{{Grindlay} {et~al.}(2001){Grindlay}, {Heinke}, {Edmonds}, \&
  {Murray}}]{grindlay01}
{Grindlay}, J.~E., {Heinke}, C., {Edmonds}, P.~D., \& {Murray}, S.~S. 2001,
  Science, 292, 2290

\bibitem[{{Harding} \& {Muslimov}(2001)}]{harding01}
{Harding}, A.~K. \& {Muslimov}, A.~G. 2001, \apj, 556, 987

\bibitem[{{Harding} \& {Muslimov}(2002)}]{harding02}
---. 2002, \apj, 568, 862

\bibitem[{{Heinke} {et~al.}(2005){Heinke}, {Grindlay}, {Edmonds}, {Cohn},
  {Lugger}, {Camilo}, {Bogdanov}, \& {Freire}}]{heinke05}
{Heinke}, C.~O., {Grindlay}, J.~E., {Edmonds}, P.~D., {Cohn}, H.~N., {Lugger},
  P.~M., {Camilo}, F., {Bogdanov}, S., \& {Freire}, P.~C. 2005, \apj, 625, 796

\bibitem[{{Houck} \& {Denicola}(2000)}]{houck00}
{Houck}, J.~C. \& {Denicola}, L.~A. 2000, in ASP Conf. Ser. 216: Astronomical
  Data Analysis Software and Systems IX, ed. N.~{Manset}, C.~{Veillet}, \&
  D.~{Crabtree}, 591

\bibitem[Lange et al.(2001)]{lange01} Lange, C., Camilo, F., 
Wex, N., Kramer, M., Backer, D.~C., Lyne, A.~G., \& Doroshenko, O.\ 2001, 
\mnras, 326, 274 

\bibitem[Lommen et al.(2000)]{lommen00} Lommen, A.~N., Zepka, 
A., Backer, D.~C., McLaughlin, M., Cordes, J.~M., Arzoumanian, Z., \& 
Xilouris, K.\ 2000, \apj, 545, 1007 

\bibitem[{{Murray} {et~al.}(1998){Murray}, {Chappell}, {Kenter}, {Kraft},
  {Meehan}, \& {Zombeck}}]{hrc2}
{Murray}, S.~S., {Chappell}, J.~H., {Kenter}, A.~T., {Kraft}, R.~P., {Meehan},
  G.~R., \& {Zombeck}, M.~V. 1998, in Proc. SPIE Vol. 3356, Space Telescopes
  and Instruments V, Pierre Y. Bely; James B. Breckinridge; Eds., 974

\bibitem[Navarro et al.(1995)]{navarro95} Navarro, J., de Bruyn, 
A.~G., Frail, D.~A., Kulkarni, S.~R., \& Lyne, A.~G.\ 1995, \apjl, 455, L55 

\bibitem[{{Nicastro} {et~al.}(2004){Nicastro}, {Cusumano}, {L{\"o}hmer},
  {Kramer}, {Kuiper}, {Hermsen}, {Mineo}, \& {Becker}}]{nicastro04}
{Nicastro}, L., {Cusumano}, G., {L{\"o}hmer}, O., {Kramer}, M., {Kuiper}, L.,
  {Hermsen}, W., {Mineo}, T., \& {Becker}, W. 2004, \aap, 413, 1065

\bibitem[Nice et al.(2005)]{nice05} Nice, D.~J., Splaver, 
E.~M., Stairs, I.~H., L{\"o}hmer, O., Jessner, A., Kramer, M., \& Cordes, 
J.~M.\ 2005, \apj, 634, 1242 

\bibitem[{{Ransom} {et~al.}(2004){Ransom}, {Stairs}, {Backer}, {Greenhill},
  {Bassa}, {Hessels}, \& {Kaspi}}]{ransom04}
{Ransom}, S.~M., {Stairs}, I.~H., {Backer}, D.~C., {Greenhill}, L.~J., {Bassa},
  C.~G., {Hessels}, J.~W.~T., \& {Kaspi}, V.~M. 2004, \apj, 604, 328

\bibitem[{{Rutledge} {et~al.}(2004){Rutledge}, {Fox}, {Kulkarni}, {Jacoby},
  {Cognard}, {Backer}, \& {Murray}}]{rutledge04}
{Rutledge}, R.~E., {Fox}, D.~W., {Kulkarni}, S.~R., {Jacoby}, B.~A., {Cognard},
  I., {Backer}, D.~C., \& {Murray}, S.~S. 2004, \apj, 613, 522

\bibitem[{{Scargle}(1998)}]{scargle98}
{Scargle}, J.~D. 1998, \apj, 504, 405

\bibitem[{{Swanepoel} {et~al.}(1996){Swanepoel}, {de Beer}, \&
  {Loots}}]{swanepoel96}
{Swanepoel}, J.~W.~H., {de Beer}, C.~F., \& {Loots}, H. 1996, \apj, 467, 261

\bibitem[{{Takahashi} {et~al.}(2001){Takahashi}, {Shibata}, {Torii}, {Saito},
  {Kawai}, {Hirayama}, {Dotani}, {Gunji}, {Sakurai}, {Stairs}, \&
  {Manchester}}]{takahashi01}
{Takahashi}, M., {Shibata}, S., {Torii}, K., {Saito}, Y., {Kawai}, N.,
  {Hirayama}, M., {Dotani}, T., {Gunji}, S., {Sakurai}, H., {Stairs}, I.~H., \&
  {Manchester}, R.~N. 2001, \apj, 554, 316

\bibitem[{{Tennant} {et~al.}(2001){Tennant}, {Becker}, {Juda}, {Elsner},
  {Kolodziejczak}, {Murray}, {O'Dell}, {Paerels}, {Swartz}, {Shibazaki}, \&
  {Weisskopf}}]{tennant01}
{Tennant}, A.~F., {Becker}, W., {Juda}, M., {Elsner}, R.~F., {Kolodziejczak},
  J.~J., {Murray}, S.~S., {O'Dell}, S.~L., {Paerels}, F., {Swartz}, D.~A.,
  {Shibazaki}, N., \& {Weisskopf}, M.~C. 2001, \apjl, 554, L173

\bibitem[{{Thorsett} {et~al.}(1999){Thorsett}, {Arzoumanian}, {Camilo}, \&
  {Lyne}}]{thorsett99}
{Thorsett}, S.~E., {Arzoumanian}, Z., {Camilo}, F., \& {Lyne}, A.~G. 1999,
  \apj, 523, 763

\bibitem[Toscano et al.(1999)]{toscano99} Toscano, M., Sandhu, 
J.~S., Bailes, M., Manchester, R.~N., Britton, M.~C., Kulkarni, S.~R., 
Anderson, S.~B., \& Stappers, B.~W.\ 1999, \mnras, 307, 925 

\bibitem[{{Webb} {et~al.}(2004{\natexlab{a}}){Webb}, {Olive}, \&
  {Barret}}]{webb04a}
{Webb}, N.~A., {Olive}, J.-F., \& {Barret}, D. 2004{\natexlab{a}}, \aap, 417,
  181

\bibitem[{{Webb} {et~al.}(2004{\natexlab{b}}){Webb}, {Olive}, {Barret},
  {Kramer}, {Cognard}, \& {L{\"o}hmer}}]{webb04b}
{Webb}, N.~A., {Olive}, J.-F., {Barret}, D., {Kramer}, M., {Cognard}, I., \&
  {L{\"o}hmer}, O. 2004{\natexlab{b}}, \aap, 419, 269

\bibitem[{{Zavlin}(2006)}]{zavlin06}
{Zavlin}, V.~E. 2006, \apj, 638, 951

\bibitem[{{Zavlin} {et~al.}(2002){Zavlin}, {Pavlov}, {Sanwal}, {Manchester},
  {Tr{\"u}mper}, {Halpern}, \& {Becker}}]{zavlin02}
{Zavlin}, V.~E., {Pavlov}, G.~G., {Sanwal}, D., {Manchester}, R.~N.,
  {Tr{\"u}mper}, J., {Halpern}, J.~P., \& {Becker}, W. 2002, \apj, 569, 894

\bibitem[{{Zombeck} {et~al.}(1995){Zombeck}, {Chappell}, {Kenter}, {Moore},
  {Murray}, {Fraser}, \& {Serio}}]{hrc1}
{Zombeck}, M.~V., {Chappell}, J.~H., {Kenter}, A.~T., {Moore}, R.~W., {Murray},
  S.~S., {Fraser}, G.~W., \& {Serio}, S. 1995, in Proc. SPIE Vol. 2518, EUV,
  X-Ray, and Gamma-Ray Instrumentation for Astronomy VI, Oswald H. Siegmund;
  John V. Vallerga; Eds., 96

\end{thebibliography}

\clearpage
\begin{deluxetable}{lccccccc} 
\tabletypesize{\scriptsize}
\tablewidth{0pt}
\tablecaption{\label{tab:other} X-ray Properties of Millisecond Pulsars Outside of \tuc.}
\tablehead{\colhead{PSR} & \colhead{$P$} & \colhead{$d$} & 
\colhead{$\tau$} & \colhead{$\log\dot{E}$} & \colhead{$\log L_X$} & \colhead{$f_p$} & Refs. \\
 & \colhead{(ms)} &  \colhead{(kpc)} & \colhead {(Gyr)}  & 
\colhead{(erg\,s$^{-1}$)} & \colhead{(erg\,s$^{-1}$)} & \colhead{(\%)} & }
\startdata
\hline
B1937$+$21              & 1.56 & 3.57                 &  0.23 & 36.04  & 33.15 & 54 $\pm$ 7 &  1,2\\
B1957$+$20              & 1.61 & 2.49                 &  2.22 & 35.04  & 31.81 & $< 60$     &  1,3\\
J0218$+$4232            & 2.32 & 2.67                 &  0.48 & 35.38  & 32.54 & 59 $\pm$ 7 &  4,5\\
B1821$-$24A (M28)       & 3.05 & 5.5\tablenotemark{a} &  0.03 & 36.35  & 33.22 & 85 $\pm$ 3 &  6,7\\
\hline
J0751$+$1807            & 3.48 & 1.15                 &  7.08 & 33.86  & 30.84 & 52 $\pm$ 8\tablenotemark{b} & 8,9 \\
J0030$+$0451            & 4.87 & 0.32                 &  7.71 & 33.53  & 30.40 & 69 $\pm$ 18 & 10,11 \\
J2124$-$3358            & 4.93 & 0.25                 &  6.01 & 33.83  & 30.23 & 56 $\pm$ 14 &  1,12\\
J1012$+$5307            & 5.26 & 0.84                 &  4.86 & 33.68  & 30.38 & 77 $\pm$ 13\tablenotemark{b} & 13,9 \\
J0437$-$4715            & 5.76 & 0.14\tablenotemark{a}&  4.91 & 33.58  & 30.47 & 40 $\pm$ 2  &  14,12\\
J1024$-$0719            & 5.16 & 0.39                 &  27   & 32.93  & 29.30 & 52 $\pm$ 22 & 1,12\\
J1744$-$1134            & 4.07 & 0.36\tablenotemark{a}&  9.1  & 33.62  & 29.49\tablenotemark{c} & ---         & 1,3 \\
J0034$-$0534            & 1.88 & 0.54                 &  6.03 & 34.48  & 29.60\tablenotemark{c} & ---         & 12 \\
\hline
No~High~Time~Resolution~Imaging& \\
\hline
B1620$-$26 (M4)         & 11.08& 1.73\tablenotemark{a}&  0.26 &34.28   & 30.08\tablenotemark{c} & ---         & 15,16\\
J1740$-$5340 (NGC~6397) & 3.65 & 2.55\tablenotemark{a}&  0.34 &35.15   & 30.9\tablenotemark{c} & ---         & 17,18\\
J1911$-$6000C (NGC~6752)& 5.28 & 4.1\tablenotemark{a} &  38.1 &32.77   & 30.34\tablenotemark{c}  & ---         & 19\\
J2140$-$2310A (M30)     & 11.02& 9.0\tablenotemark{a}  &$>$0.08&$<$34.79& 30.64\tablenotemark{c} & ---         & 20\\
\enddata 
\tablenotetext{a}{Accurate distance measurement.}
\tablenotetext{b}{Detection significance is low.}
\tablenotetext{c}{X-ray luminosity in the 0.5--2.5 keV band.}
\tablecomments{ 
All distances are estimated from the pulsar dispersion measures and
the model of Galactic distribution of free electrons \citep{cordes02},
except where noted.  X-ray luminosities are quoted in the 0.2--10 keV
band as adopted from Table~1 of \cite{zavlin06} and references therein, except
where noted.  Pulsed fractions are quoted in roughly in the HRC-S band
(0.1--2.0 keV), but see references for the specific bandpass.  The spectra
of MSPs above the first horizontal line are dominated by non-thermal
X-ray emission. Those below the line have significant thermal
components or are indeterminate (and thus presumed to be thermal) in
nature. References:
1 -- \cite{toscano99},
2 -- \cite{nicastro04},
3 -- \cite{bt99},
4 -- \cite{navarro95},
5 -- \cite{webb04a},
6 -- \cite{becker03},
7 -- \cite{rutledge04},
8 -- \cite{nice05},
9 -- \cite{webb04b},
10 -- \cite{lommen00},
11 -- \cite{becker02},
12 -- \cite{zavlin06}
13 -- \cite{lange01},
14 -- \cite{zavlin02},
15 -- \cite{thorsett99},
16 -- \cite{bassa04},
17 -- \cite{damico01},
18 -- \cite{grindlay02},
19 -- \cite{damico02},
20 -- \cite{ransom04}.
}
\end{deluxetable}  

\clearpage
\begin{deluxetable}{llccc}
\tablecaption{\label{tab:obs} HRC-S Observations.}
\tablewidth{10cm} 
\tablehead{
\colhead{ObsID}  & \colhead{Obs. Start  }& \colhead{Obs. Start }& \colhead{Exp. Time}\\
 & \colhead{ (UT) }& \colhead{ (MJD) }& \colhead{ (sec)}}
\startdata
5542 & 2005 Dec 19.29 & 53723.79 & 51918.2 \\
5543 & 2005 Dec 20.62 & 53725.12 & 53962.6 \\
5544 & 2005 Dec 21.98 & 53726.48 & 52036.6 \\
5545 & 2005 Dec 23.21 & 53727.71 & 54203.5 \\
6237 & 2005 Dec 24.59 & 53729.09 & 51920.8 \\
6238 & 2005 Dec 25.88 & 53730.38 & 50887.6 \\
5546 & 2005 Dec 27.23 & 53731.73 & 51939.3 \\
6230 & 2005 Dec 28.57 & 53733.07 & 52401.0 \\
6231 & 2005 Dec 29.91 & 53734.41 & 48963.7 \\
6232 & 2005 Dec 31.22 & 53735.72 & 49139.0 \\
6233 & 2005 Jan 2.24  & 53737.74 & 103433.2\\
6235 & 2005 Jan 4.17  & 53739.67 & 51932.1 \\
6236 & 2005 Jan 5.48  & 53740.98 & 54729.8 \\
6239 & 2005 Jan 6.92  & 53742.42 & 52241.6 \\
6240 & 2005 Jan 8.10  & 53743.60 & 54178.4 \\
\enddata
\end{deluxetable}

\clearpage
\begin{deluxetable}{lccccccccccc}  
\setlength{\tabcolsep}{0.05in}     
\rotate                            
\tabletypesize{\scriptsize}        
\tablecaption{\label{tab:money} HRC-S Derived X-ray Properties of the \tuc\ Millisecond Pulsars.}
\tablehead{
\colhead{Name}&\colhead{Total\tablenotemark{a}}& \colhead{Background} & \colhead{Extraction}&\colhead{$\log (L_X)$\tablenotemark{b}}&
\colhead{Harmonics\tablenotemark{c}} & \colhead{$Z^2_n$\tablenotemark{c}}& \colhead{Significance\tablenotemark{c}} & 
\colhead{$f_{p,{\rm radio}}$\tablenotemark{c,d}} &  \colhead{$f_{p,{\rm sine}}$\tablenotemark{c,d}}& \colhead{$f_p$\tablenotemark{c}} \\
 & \colhead{Counts} &  \colhead{Counts} &\colhead{Radius} & (0.5-2.0\,keV) & \colhead{($n$)}
& & \colhead{($\sigma$)} \\
}
\startdata
47 Tuc C                     & 173 & 94.8 & 1\farcs3 & 30.12  & 1  &   2.5 &  1.5  & $<$76 &  $<100$ &  \\ 
47 Tuc D                     & 221 & 94.8 & 1\farcs3 & 30.33  & 2  &  21.7 &  3.9  & ---&  --- &  $70 \pm 21$\\
47 Tuc E\tablenotemark{g}    & 254 & 94.8 & 1\farcs3 & 30.43  & 1  &  12.5 &  3.3  & ---&  --- &  $50 \pm 19$\\
47 Tuc F\tablenotemark{e}    & 413 & 80.7 & 1\farcs2 & $<$30.75  & 1  &  10.1 &  2.9  & ---&  --- & $2 \pm 12$\\
47 Tuc G\tablenotemark{f}    & 322 & 80.7 & 1\farcs2 & $<$30.61  & 1  &   1.0 &  1.0  & $<100$ &  $<100$  &  \\
47 Tuc H\tablenotemark{g}    & 176 & 94.8 & 1\farcs3 & 30.14  & 2  &  12.7 &  2.8  & ---&  --- &  $26 \pm 20$\\
47 Tuc I\tablenotemark{f,g}  & 322 & 80.7 & 1\farcs2 & $<$30.61  & 1  &   6.0 &  2.3  & $<$81 &  $<100$  &  \\
47 Tuc J\tablenotemark{g}    & 266 & 94.8 & 1\farcs3 & 30.46  & 1  &   6.5 &  2.3  & $<$38 &  $<$77  &  \\
47 Tuc L                     & 342 & 35.9 & 0\farcs8 & 30.71  & 2  &  10.0 &  2.4  & $<$50 & $<$ 49  &  \\
47 Tuc M                     & 151 & 94.8 & 1\farcs3 & 29.98  & 1  &   2.2 &  1.4  & $<100$ &  $<100$  &  \\
47 Tuc N                     & 186 & 94.8 & 1\farcs3 & 30.19  & 1  &   4.8 &  2.0  & $<$73 &  $<100$  &  \\
47 Tuc O\tablenotemark{g}    & 431 & 94.1\tablenotemark{h} & 1\farcs2 & 30.77  & 3  &  39.1 &  5.1  & ---&  --- & $81 \pm 21$ \\
47 Tuc Q\tablenotemark{g}    & 186 & 94.8 & 1\farcs3 & 30.19  & 6  &  28.3 &  3.1  & ---&  --- &  $83 \pm 42$\\
47 Tuc R\tablenotemark{g}    & 288 & 80.4\tablenotemark{h} & 1\farcs1 & 30.57  & 2  &  24.1 &  4.1  & ---&  --- & $63 \pm 29$  \\
47 Tuc S\tablenotemark{e,g}  & 413 & 80.7 & 1\farcs2 & $<$30.75  & 8  &  33.4 &  3.1  & ---&  --- &$20 \pm 15$\\
47 Tuc T\tablenotemark{g}    & 133 & 80.7 & 1\farcs2 & 29.94  & 1  &   1.2 &  1.1  & $<100$ &  $<100$  &  \\
47 Tuc U\tablenotemark{g}    & 193 & 94.8 & 1\farcs3 & 30.22  & 1  &   0.1 &  0.7  & $<100$ &  $<100$  &  \\
47 Tuc W\tablenotemark{g}    & 433 & 80.7 & 1\farcs2 & 30.77  & 1  &   0.8 &  1.0  & $<$48 &  $<$48  &  \\
47 Tuc Y\tablenotemark{g}    & 218 & 94.8 & 1\farcs3 & 30.32  & 1  &   6.5 &  2.3  & $<$40 &  $<$96  &  \\
\enddata
\tablenotetext{a}{ This number includes the expect background counts listed in the subsequent column.}
\tablenotetext{b}{ X-ray luminosity (logarithm of \cgslum) derived in the band 0.5--2.0\,keV.}
\tablenotetext{c}{ See the text (\S~\ref{sec:rot}) for a description of these parameters.}
\tablenotetext{d}{ $<$100\% means the undetected pulsation is
consistent with 100\% pulsed signal, and therefore is unconstrained by
simulations.}
\tablenotetext{e}{ \tuc~F and S have overlapping positions. The total counts represent all
photons extracted from the 1\farcs2 extraction radius and the background counts are only
those expected from a uniform background.}
\tablenotetext{f}{ \tuc~G and I have overlapping positions. The total counts represent all
photons extracted from the 1\farcs2 extraction radius and the background counts are only
those expected from a uniform background.}
\tablenotetext{g}{ Binary MSP.}
\tablenotetext{h}{ Includes an estimate of the contamination due to source crowding (see \S~\ref{sec:rot}).}
\end{deluxetable}    

\clearpage
\begin{deluxetable}{cccc}
\tablecaption{\label{tab:prob} Summary of \tuc\ MSP Detection Significance.}
\tablewidth{12cm} 
\tablehead{
\colhead{Pulsation Detection}& \colhead{Number of} & \colhead{Binomial}   & \colhead{False Detection}\\
\colhead{Threshold}          & \colhead{Detections}& \colhead{Probability}& \colhead{Probability} \\
\colhead{(1)}  & \colhead{(2)} & \colhead{(3)} & \colhead{(4)}
}
\startdata
99\%      &   8  &  6.9\tee{-12}  & 17.5\% \\
99.73\%   &   6  &  1.0\tee{-11}  & 5\% \\
99.947\%  &   3  &  1.4\tee{-7}   & 1\% \\
\enddata
\tablecomments{
(1) -- Confidence level of $Z^2_n$ above which a \tuc\ MSP is labeled `variable.'
(2) -- The number of `variable' MSPs for the given confidence level.
(3) -- The binomial probability that the number of `variable' MSPs is due to chance.
(4) -- The probability that one of the `variable' MSPs is a statistical fluctuation.
}
\end{deluxetable}

\clearpage
\thispagestyle{empty}
\setlength{\voffset}{-15mm}

\clearpage

\clearpage

\end{document}